\documentclass[aps,twocolumn,showpacs,preprintnumbers,amsmath,amssymb,superscriptaddress]{revtex4-1}

\usepackage{hyperref}

\usepackage{amssymb}
\usepackage{color}

\def \Prague{Institute of Physics, Charles University in Prague, Faculty of Mathematics and Physics, Ke Karlovu 5, Prague 2, CZ-121~16, Czech Republic}

\usepackage{epsfig}
\usepackage{amsmath}

\begin{document}
\title{Efficient Charge Collection in Coplanar Grid Radiation Detectors}
\author{J. \surname{Kunc}} \affiliation{\Prague}

\email{kunc@karlov.mff.cuni.cz}
\author{P. \surname{Praus}}
\author{E. \surname{Belas}} \affiliation{\Prague}
\author{V. \surname{D\v{e}di\v{c}}} \affiliation{\Prague}
\author{J. \surname{Pek\'{a}rek}} \affiliation{\Prague}
\author{R. \surname{Grill}} \affiliation{\Prague}

\date{\today}

\begin{abstract}
We have modeled laser-induced transient current waveforms in radiation coplanar grid detectors. Poisson's equation has been solved by finite element method and currents induced by photo-generated charge were obtained using Shockley-Ramo theorem. The spectral response on a radiation flux has been modeled by Monte-Carlo simulations. We show 10$\times$ improved spectral resolution of coplanar grid detector using differential signal sensing. We model the current waveform dependence on doping, depletion width, diffusion and detector shielding and their mutual dependence is discussed in terms of detector optimization. The numerical simulations are successfully compared to experimental data and further model simplifications are proposed. The space charge below electrodes and a non-homogeneous electric field on a coplanar grid anode are found to be the dominant contributions to laser-induced transient current waveforms.
\end{abstract}


\maketitle
\section{Introduction}
High energy radiation can be detected directly by converting photons into electrical signals, or indirectly using scintilators and photodiodes. Direct conversion detectors are expected to have higher signal-to-noise ration, higher spatial and temporal resolution, even on a single photon counting level. This could lead to spectrally resolved computed tomography and advanced medical X-ray imaging~\cite{Overdick2009}. An efficient detector optimization is a challenging technological task. The single crystal growth is one of them, high mobility-relaxation time product ($\mu\tau$) is another one and least but not the last one is different mobility and trapping cross-section of photo-generated holes. The latter one contributes to significant degradation of temporal resolution. A wide range of materials~\cite{Nikl2006,KasapSensors2011,Overdick2009,HampaiLiF2011} provide good sensitivity, but poor temporal performance is achieved. The low mobility of holes, compared to electron mobility, have been solved by coplanar grid (CPG) detection~\cite{Luke1995} in CdZnTe detectors, semiconductor analogue of Frisch grids in gas ion chambers. 

As the technology of majority of proposed materials needs yet to mature, the choice of GaAs, crystalline CdTe or CdZnTe is mandatory for counting and high energy resolution applications. The problem with different electron and hole mobilities is overcome using the coplanar grid electrodes and it has been proven also to be efficient way to increase specral resolution~\cite{ref1}. The coplanar-grid as well as other electron-only detection techniques~\cite{Zhang2013,Owens2007} are effective in overcoming some of the material problems of CdZnTe and, consequently, have led to efficient gamma-ray detectors with good energy resolution while operating at room temperature~\cite{AmmanJAP2002}.

The main limitations yet remain to be high quality crystalline material and polarization effects. Their reduction has been addressed both experimentally and theoretically. Atomistic simulations~\cite{WardPRB2012,WardPRB85-2012} are a powerful method for exploring crystalline defects at a resolution unattainable by experimental techniques~\cite{refColaAPL}. Determination of electric field shape precisely due to relatively complicated electrodes structure and application of the bias and intergrid voltage simultaneously is experimentally also difficult task.
Though exact solutions are often tedious, simplified models have been also proposed. A small part of pixel detector has been considered~\cite{Guerra2009} and spectral response modeled. Polarization and carrier trapping was solved by simplified set of kinetic equations~\cite{BalePRB2008} in CdZnTe detectors, the effect of transport parameters' inhomogenities (charge density, mobility, potential fluctuations) and hot carriers were discussed~\cite{HarrisonPRB2008} and plasmons in solid state radiation detectors were studied~\cite{ChoyEPL1993}. The Laser induced Transient Current Technique (L---TCT) provides information about charge collection dynamics and detector polarization~\cite{ref2,ref3,ref4,ref5}. It has been measured on pixel~\cite{Overdick2009} and coplanar grid detectors~\cite{PrausAPL2016}. Though it is important for further detector optimization, the theoretical treatment is usually given only in simplified experimental geometries and basic formulation of Schockley-Ramo theorem~\cite{Prettyman1998}. The coplanar grid detectors have been modeled also only in simplified geometry and spectral response has been calculated only~\cite{McGregor1998,Luke1995,Prettyman1998,KozerovAPL2007}. Sparse realistic detector geometries were considered for X-ray spectra simulations~\cite{Ma2014,Huang2015}, however, no temporal response has been calculated.

We provide here an insight on carrier transport in a radiation coplanar grid detector based on CdZnTe, valid also for other semiconductors used in direct conversion detectors. We determine  influence of bulk semiconductor parameters on a temporal and spectral resolution in real detector geometry. The requirement on efficient charge collection is discussed with respect to effects of depletion width, diffusion and detector shielding. The L---TCT waveforms compared to numerical model allow evaluation of the electric field profile inside coplanar grid detector with high spatial resolution unattainable experimentally. The spectral resolution improvement is quantitatively determined in a coplanar grid detector and the differential current sensing~\cite{PrettymanDiff1999,Luke1995,McGregor1998} is shown to overcome problems of low hole mobility, high hole trapping cross-section and random nature of X-/gamma ray photon absorption within a detector. We also aim to show simplified analytically based considerations to understand detector operation, which is useful for qualitative optimization without need of elaborate numerical calculations. 

\section{Methods}
The L---TCT response has been modeled by drift transport mechanism of photo-generated carriers. The above band gap photo-generation is assumed. Holes have been neglected for their small mobility $\mu_h\cong 100$~cm$^2$V$^{-1}$s$^{-1}$ and short life time~\cite{Fink2006}. For this reason we assume electrons only~\cite{Luke1995} and collection times much shorter than electron life time. In L---TCT, holes are swept to cathode in subnanosecond times, hence, their influence is negligible. The current response caused by moving charge between electrodes have been calculated using Shockley-Ramo theorem. The foundation of Shockley-Ramo theorem is based on known electrostatic potential in two cases. The first one consists of real electrostatic potential distribution within a detector. The second is a potential distribution of a weighting potential. Here, the electrode of interest is biased to 1~V and all other electrodes are at 0~V. The corresponding electric  field determines the charge response on a given electrode and the time derivative determines the current response~\cite{ref9,ref8,ref10,ref11}. The full real potential and corresponding electric field determines the trajectory of photo-generated carriers. We assume only small enough concentration of photo-generated carriers that they do not spread by diffusion or electrostatic repulsion on the length scale of detector size before they are all swept by anode. This is clearly well fulfilled since the diffusion time broadening is about 15~ns and carriers move about 1~$\mu$s from cathode to anode. The limitations of this assumption are discussed comparing the model and experimental data.

The L---TCT current response is modeled by assuming 1.4~mm wide laser spot size on partially transparent cathode. We trace 300 electron trajectories equally spaced on a cathode center. The motion of each electron (electron packet) has been calculated by integrating their drift velocity. The current response is then calculated for each of 300 trajectories and summed up. The resulting summed current response has been compared to experimental data. We also assume positively charged depletion width below cathode and negatively charged space region below anode, as commonly observed in CPG. These space charge regions give rise to current decrease (increase) at the time of 0 (1~$\mu$s). 

The experiments done by hard X-ray and $\gamma$-ray transient current technique (X---TCT) has been modeled by solution of Poisson equation and, in the second computation step, electron trajectories have been obtained by integrating their velocities (as in L---TCT curves). However, here, the initial position of photo-generated carriers have been determined in a Monte-Carlo loop. Random position of photo-generated carriers is a model of large absorption length of X/gamma-ray radiation, hence the model holds for more than 100's~eV energetic photons. Here, we have not summed all trajectories with a random initial position. Instead, the current response has been calculated for each trajectory separately and the total collected charge has been calculated by integrating current response in time. The electron current response has been considered only since holes' mobility is $10\times$ lower~\cite{Fink2006} and holes have higher trapping cross-section. We have traced 1500 trajectories and total collected charge from electron motion has been statistically analyzed. 

The electrostatic potential has been calculated by solving Poisson's equation $-\Delta\varphi=\frac{\rho}{\epsilon}$ by finite element method. The detector geometry has been chosen to properly model a Redlen Inc. state-of-the-art CdZnTe CPG detector. The mesh triangulation has been used to properly describe size effects of rectangular metal anode grid. The anode behaves as a set of parallel charged wires when charge is far from the anode compared to anode characteristic size (anode strip width). And, the anode behaves like infinitely large planar electrode if the charge is close enough to anode compared to characteristic anode size. We consider the latter case as a measure of sufficiently dense triangulation. This regime can be analytically treated as well, and pronounces itself as a constant electric field and linear-in-distance electrostatic potential.

The strong electric field at both electrodes causes velocity saturation. The field maximal values are $\approx 10^{4}$~V/cm, giving drift speed up to $1\times10^7$~cm/s assuming electron mobility is $\mu\approx 1000~$cm$^{2}$V$^{-1}$s$^{-1}$. The saturation velocity in CdZnTe is about $v_{sat}\approx1\times10^7$~cm/s~\cite{refRuzin} therefore the velocity field dependence is expected to show non-linear behavior. The velocity saturation~\cite{refSze} is taken into account by
$v=\frac{\mu E_{loc}}{\sqrt{1+\left(\frac{\mu E_{loc}}{v_{sat}}\right)^2}}$.

\section{Samples and experimental setup}
The modeled L---TCT waveforms have been compared to experimental data measured on coplanar grid CdZnTe detector. The detector, purchased from Redlen Inc., has dimensions $10\times10\times10$~mm$^3$. The details of our L--–TCT setup~\cite{ref6,ref7} and data proccessing are descibed elswhere~\cite{PrausAPL2016}. The detector cathode was irradiated by optical pulses ($\approx3$~ns as FWHM, 200~ps rise and fall time) using the laser diode (660~nm, 300~mW) that is powered by an ultra-fast pulse generator. Pulse output energy was 0.4~nJ and repetition rate was 100~Hz. A collimation and imaging optics and a translation stage is used in the setup to adjust the laser diode beam onto the examined spot of $\approx3$~mm$^2$ on the cathode. Neutral density disc filter is placed in the beam line for optimal intensity attenuation of the laser pulse. Negative bias $V_B$ is applied to the CPG detector cathode (up to ­1700~V) and supplementary intergrid voltage $V_{IG}$ between the collecting and non–collecting anode grid is introduced by an additional adjustable low noise voltage supply ($0~-~200$~V). 

\begin{figure}[t!]
\centering
\includegraphics[width=9cm]{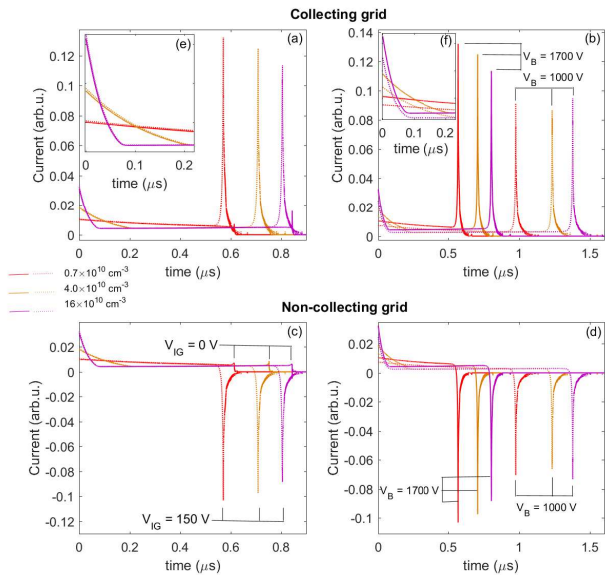}
\caption{Doping dependence of L---TCT response curves. Doping densities $N_D=0.7$, 4.0 and 16.0 $\times10^{10}$~cm$^{-3}$ are distinguished by red, orange and violet color. Solid (dashed) curves in (a,c,e) show numerical simulation for intergrid bias $V_{IG}=0$~V (150~V). Solid (dashed) curves in (b,d,f) show numerical simulation for negative cathode bias $V_B=1700$~V (1000~V). The insets (e,f) depict detail of the current response around $t\approx0~\mu$s. The space charge width below cathode is $0.64DW$. The simulation is shown for (a,b,e,f) collecting grid  and (c,d) non-collecting grid. The I$_G$ trend (a,c,e) is studied for negative V$_B$=1700~V and the V$_B$ trend (b,d,f) is studied for V$_{IG}$=150~V.}
\label{Fig1}
\end{figure} 
\section{Charged impurities, depletion width and diffusion}
The effect of doping is shown  in Fig.~\ref{Fig1} for positive space charge $N_D=0.7$, 4.0 and 16$\times10^{10}$~cm$^{-3}$. The current response in L---TCT is modeled for collecting grid (CG) in Fig.~\ref{Fig1}~(a,b,e,f) and for non-collecting grid (NCG) in Fig.~\ref{Fig1}~(c,d). The current response can be split into three characteristic intervals. The first one is given by charge deceleration in a positive space charge below cathode. The second regime is at intermediate times when the charge moves through the part of the detector without any fixed space charge. The third regime is a time of charge collection at the anode. It can be seen from numerical simulation that the peak current response at the collection time increases with decreasing doping. The initial current peak is decreasing with decreasing doping and ideally disappears. The initial current peak can give rise to a spurious signal contributing to a low energy spectral shoulder in $\gamma$-ray spectra. The anode intergrid bias is shown for two values 0 and 150 V in Fig.~\ref{Fig1}~(a) and (c). We note that the intergrid bias does not influence the initial current response, as is shown in the inset Fig.~\ref{Fig1}~(e), in contrast to cathode bias, inset (f). This is caused by the local effects of intergrid bias on a small area around anode. The cathode bias $V_B$ (cathode to NCG bias) influences mean field within a detector, thus changing current response for all times from the photo-carrier generation to their complete collection on CG.

\begin{figure}[t!]
\centering
\includegraphics[width=9cm]{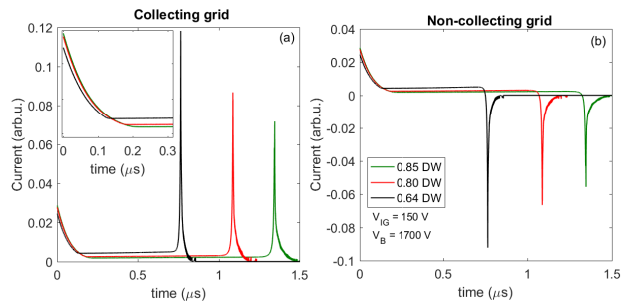}
\caption{Space charge width dependence of L---TCT response curves. The space charge width is measured in units of equilibrium depletion width DW for $N_D=8\times10^{10}$~cm$^{-3}$. The black, red and green curves correspond to depletion width 0.64, 0.80 and 0.85~DW, respectively. The numerical simulation is shown for (a) collecting and (b) non-collecting grid. The inset in (a) depicts detail of current evolution at $t\approx0~\mu$s. The intergrid bias $V_{IG}=150$~V and cathode bias negative $V_{B}=1700$~V.}
\label{Fig2}
\end{figure} 
The cathode depletion width dependence of the L---TCT waveform is shown in Fig.~\ref{Fig2}. The cathode is a planar electrode and the constant current is expected for uncharged bulk semiconductor. The space charge (fixed or mobile) causes deceleration of carriers hence decreasing current. The equilibrium depletion width ($DW$) is given by $DW=\sqrt{\frac{2\epsilon V}{qN_D}}$ , where $N_D$ is doping density (or fixed space charge), $V$ is a voltage drop within a depletion width and $\epsilon$ is permitivity. The width of the depletion region varies from this simple model when carrier trapping and de-trapping is assumed, or more than one trapping level is located in the vicinity of the Fermi level. It can be seen from Fig.~\ref{Fig2} that the peak current increases with decreasing depletion width. This effect is due to the reduced total voltage drop in a depletion region and consequently higher voltage drop at the anode. The latter causing higher electric field and higher current. We note here, that this is in contrast to doping density. The current response of a NCG is shown in Fig.~(2)~(b). The current response is, similarly to CG, stronger for thinner depletion width.

\begin{figure}[t!]
\centering
\includegraphics[width=9cm]{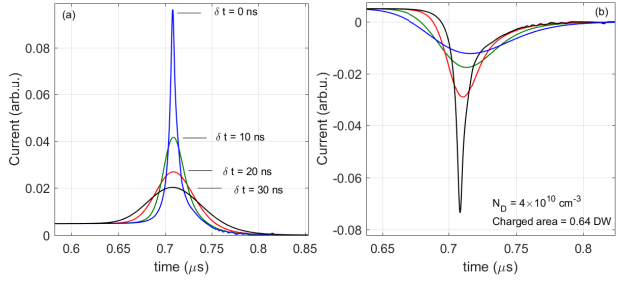}
\caption{The effect of diffusion on L---TCT current response; positive space charge below cathode only (0.64DW, $N_D=4\times10^{10}$~cm$^{-3}$). Diffusion peak broadening $\delta t=$~0, 10, 20 and 30~ns has been assumed. The effect of diffusion is simulated for (a) collecting and (b) non-collecting grid. The intergrid bias $V_{IG}=150$~V and cathode bias $V_{B}=1700$~V.}
\label{Fig3}
\end{figure} 
The effect of diffusion is shown in Fig.~\ref{Fig3}. Diffusion takes important role for long enough transient times $t$ when $\delta$t is comparable or larger than drift current temporal width. For this reason we show only the current peak modification at the charge collection on anode. The temporal broadening can be estimated from $\delta t=\sqrt{2D_et}/v$, where $D_e=\mu_e\frac{k_BT}{e}$ is electron diffusion constant related to electron mobility $\mu_e$, temperature $T$ and mean drift velocity $v$, which can be estimated from the transient time $t$ and detector height $d$; $v=d/t$. The current response for CG and NCG is shown for four diffusion broadening times (0, 10, 20, 30~ns). The current response is calculated for $N_D=4\times10^{10}$~cm$^{-3}$ and space charge width 0.64 of the equilibrium depletion width.

\section{Electrostatic shielding}
\begin{figure}[t!]
\centering
\includegraphics[width=9cm]{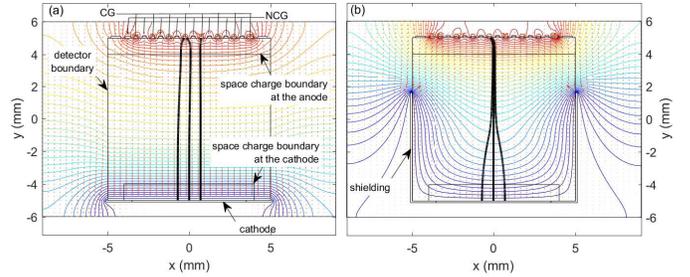}
\caption{(color online) Equipotential lines in (a) unshielded and (b) shielded detector. A detector area is depicted by thin solid black lines. The cathode is at 0~V (bottom planar contact), CG anode at negative 1850~V and NCG anode at negative 1700~V (coplanar grid anode at the top). Three representative electron trajectories are depicted as thick black lines. }
\label{Fig9}
\end{figure}

\begin{figure}[t!]
\centering
\includegraphics[width=7cm]{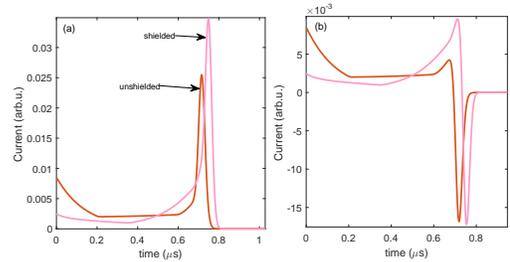}
\caption{The current waveform on (a) CG and (b) NCG anode for (orange) unshielded and (light red) shielded detector. Calculated using potential spatial distribution shown in Fig.~\ref{Fig9}.}
\label{Fig8}
\end{figure}

We study here the effect of electrostatic shielding on inner potential distribution within a detector. The detector shielding has been assumed to cover 2/3 of detector height. This coverage maximizes the shielding effect yet reduces a risk of material and surrounding air dielectric breakdown caused by strong electric field between cathode (top shielding edge) and anode. The critical electric field $E_c=3$~kV/mm in air provides possibility to apply $V_{B,max}\approx9$~kV between shielded cathode and anode (shielding-anode gap 3~mm for 2/3 of detector height coverage). We show three trial electron trajectories in Fig.~\ref{Fig9} for (a) unshielded and (b) shielded detector. The electron trajectories are almost parallel for unshielded detector and they tend to get localized to one trajectory in the shielded case.  
The current response is shown in Fig.~\ref{Fig8}. The initial current peak decreases to about 40\% of the unshielded value and the collection peak current rises by $\approx30\%$. Both effects improve detector functionality. The first peak reduction will lead to lower spurious signal and the stronger collection peak to better signal-to-noise ratio.

\section{Experimental data}
\begin{figure}[t!]
\centering
\includegraphics[width=7cm]{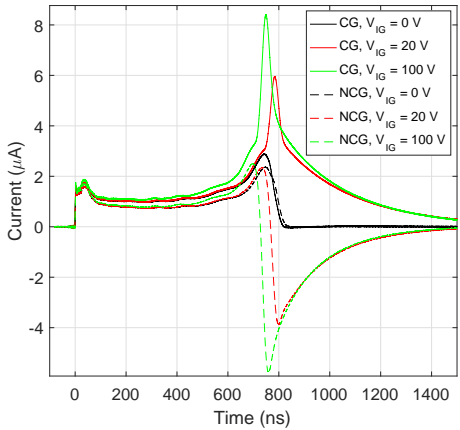}
\caption{Experimental data of L---TCT measured on a collecting grid (solid curves) and non-collecting grid (dashed curves) of non-shielded detector for negative $V_B=1500$~V and three inter-grid biases $V_{IG}=0$~V, 20~V, 100~V depicted by black, red and green curves, respectively.}
\label{Fig7}
\end{figure}

The measured L---TCT waveforms in a coplanar grid detector on a CG and NCG anode are shown in Fig.~\ref{Fig7}. The data show good qualitative agreement, see Fig.~\ref{Fig8}, and in many aspects also quantitative comparison with simulated data can be made, especially concerning relative intensities of current peaks at the beginning and at the end of charge collection. The initially decreasing current is caused by decelerating electrons in a positive space charge below the cathode. The second collection current peak is stronger than the first one especially due to the strong electric field between CG and NCG. This current originates also in a negative space charge below anode. The current peak at the time of charge collection $t\cong700-800$~ns is broadened by electron diffusion. The amount of broadening estimated from the numerical simulation is $\delta t\approx$10~ns. 

\section{Differential current sensing}
\begin{figure}[t!]
\centering
\includegraphics[width=6cm]{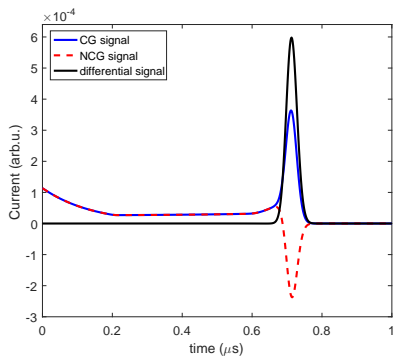}
\caption{L---TCT current response for $V_B/3$ voltage drop at the cathode (positive space charge $N_D=4\times10^{10}$~cm$^{-3}$) and $V_B/10$ voltage drop at the anode (negative space charge $N_A=5\times10^{10}$~cm$^{-3}$). The current response measured at the collecting grid (solid blue curve), non-collecting grid (dashed red curve) and differential signal (solid black curve). The applied biases are $V_{IG}=150$~V and negative $V_B=1700$~V.}
\label{Fig5}
\end{figure}

\begin{figure}[t!]
\centering
\includegraphics[width=9cm]{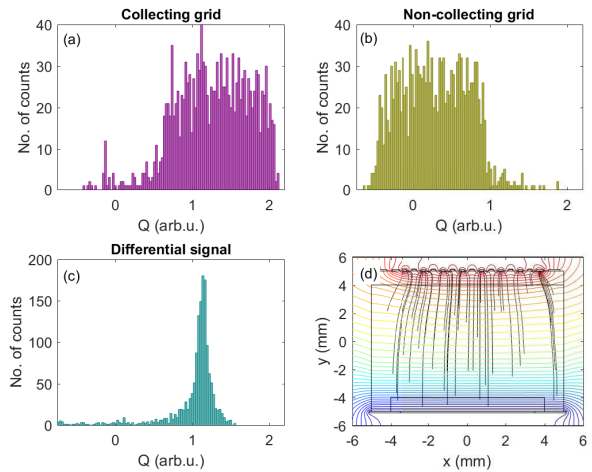}
\caption{(color online) Monte-Carlo simulation of 1500 single $\gamma$-ray photon absorption events. Doping density, depletion widths and applied biases are the same as in Fig.~\ref{Fig5}. The absorption probability is assumed to be equal throughout the detector volume. Currents induced by holes are neglected. The histograms (a,b,c) depict statistics of collected charge for all 1500 absorption events. The statistics is shown for the case of current measured on (a) collecting, (b) non-collecting grid and (c) differential signal between collecting and non-collecting grid. Equipotential lines (blue to red contour lines), detector area (solid black lines) and 50 selected random absorption events are shown in (d). }
\label{Fig6}
\end{figure}

Here we study importance of the measurement technique to detect photo-generated carriers. An example of three current waveforms detected on CG, NCG and differentially measured signal between CG and NCG is shown in Fig.~\ref{Fig5}. The advantage of differential sensing is lack of any signal before charge reaches coplanar grid anode. There is no influence of the positive space charge at cathode, since such a current response is equal for CG and NCG. We note that the differential sensing does not bring significant improvement for L---TCT, where the differential current waveform can be formed by data post-processing, as has been done in numerical simulation in Fig.~\ref{Fig5} to illustrate the method. The advantage is more pronounced when $\gamma$-ray photons are detected. Here, the absorption events occur randomly within a detector volume and they contribute to signal broadening. The $\gamma$-ray photon detection is modeled in Fig.~\ref{Fig6}.

We have traced 1500 trajectories of randomly generated photo-excited electrons and the total electron-induced charge has been calculated. An example of 50 random trajectories is shown in Fig.~\ref{Fig6}~(d). We note, that if holes are considered the total collected charge is $1e$, however, we assume low hole mobility and high trapping cross-section for holes. This leads to much weaker current response, or, the current response on very long time scales in comparison to electron induced TCT waveforms. For this reason, electron traced from below cathode gives larger induced charge than electrons traced from the anode vicinity. The statistics of all Monte-Carlo simulated absorption events are shown in Fig.~\ref{Fig6}~(a,b,c) for CG, NCG and for differential sensing, respectively. The CG and NCG current waveforms have rectangular like shape caused by equal distribution of absorption events within a detector. The mean collected charge on NCG is shifted towards zero because ideally no charge is collected by NCG. We consider here negative space charge at the anode. This space charge screens the electric field of neutral bulk detector (no charge collected by NCG) and it allows certain electron trajectories to be collected by NCG. 

In contrast to CG current response, the differential sensing, see Fig.~\ref{Fig6}~(c) shows the same mean charge collection, however, the Full Width at Half Maximum is 10$\times$ smaller. This is caused by the signal insensitivity to the position of random-in-nature absorption events of $\gamma$-ray photons. We also note that there is no spurious signal at 0~ns, Fig.~\ref{Fig5}, caused by positive space charge hence reduced low energy spectral shoulder~\cite{GiazEPL2015} is expected beside the Compton scattering contribution. The effect of holes is also reduced regardless of assumptions made in our simulations. Holes are attracted towards cathode giving even weaker response on a CG anode. 

\section{Discussion}
General considerations to understand current waveforms follow from Shochley-Ramo theorem $i=eE_vv$. Decreasing/increasing current shows on decreasing/increasing electric field in a positive/negative space charge at cathode/anode. The space charge at anode forms weak current response at $V_{IG}=$~0V and it causes current shoulder for early time of charge collection. The dominant signal at the charge collection time for $V_{IG}>$~0 is caused by accelerated electrons in a spatially modulated electric field by inter-grid bias $V_{IG}$. 
More specifically, we have shown that the cathode bias increases the peak current response. It also increases the initial signal caused by positive space charge at the cathode~\ref{Fig1}~(b). This can lead to spurious signal leading to low  energy spectral shoulder in X ray spectra. At the same time, since the charge is swept faster through detector bulk, there is lower probability for carrier trapping and detector polarization. The lowest doping density is still the ultimate option here because it worsens the charge collection efficiency.
The effect of diffusion is twofold. First, the higher is diffusion/mobility, the faster is a response time at a given detector size. Second, as absorption efficiency increases with detector size, diffusion lowers the signal. Hence, for fast response and strong current signal, thinner detector is acceptable. We also point out that thinner detector volume also eliminates carrier trapping. 
The optimal detector thickness and coplanar grid dimensions can be established for differential signal sensing. The final peak width in the case of differential sensing is caused by space charge at the anode, grid size (signal increases for denser anode grid) and by carrier diffusion. Low diffusion rate also leads to better signal-to-noise ratio. The diffusion length on a distance of detector size should be smaller than anode intergrid spacing. If diffusion length is comparable or larger than the anode intergrid spacing, the CG and NCG diffusion dominated current will be equal, and, differential signal will be negligible. 
The shielding has been shown to improve the current waveforms in a solely drift model, as shown experimentally~\cite{JoCAP2015}. It improves signal to noise ratio, deceases the initial spurious signal caused by positive space charge at the cathode and eliminates noise from the external detector environment. Contrary to that, it slows the detector response time by about $5\%$ and directs carriers into one single conducting channel inside detector. The channeling effect leads to increased diffusion and electron repulsion even for intermediate times before charge collection. 

Next, further simplifications of the proposed numerical model are discussed. It is shown that many current waveform characteristics can be explained separately one by one using simple physical considerations. These are based on electrostatic potential spatial distribution of the space charge in bulk semiconductor and assuming proper scaling of charge-to-electrode distance. If charge is close to planar cathode with respect to the cathode dimension, the electric field can be considered constant and electrostatic potential linearly scales with distance from cathode. In the presence of space charge below cathode and charge-cathode proximity, the field scales linearly and electrostatic potential quadratically with the charge-cathode distance. The Schockley-Ramo theorem than gives quadratic current waveform. The situation at the anode has to be split into two regimes. When electron is far from anode compared to the anode dimension, the electric field can be tought as that of charged wire. The intergrid potential difference will be negligible and the electron effectively moves in the effective electric field given by mean potential on CG and NCG anode. This mean potential influences the transient time by the order of 100D/L~\%, where D is anode width and L is a detector thickness. When electron approaches anode on a distance much smaller than anode width, the situation becomes equal to the one at cathode in close charge-cathode proximity. The iterative schemes can be applied to describe analytically field profiles through whole detector volume. Deviations from these assumptions lead to another model parameters. We have studied negative space charge below anode as one example and discussed diffusion due to localized carriers in one narrow potential minimum when shielding is applied as a second example of these deviations. The first additional parameter is easily taken into account in our drift model, the second one exceeds our model limitations. The proper treatment of the latter case has to be solved by coupled set of drift-diffusion, continuity and Poisson equation. The numerical hardness of such simulation however hinders basic physical considerations as discussed above. 

\section{Conclusion}
We have provided simple drift-based model to understand laser induced transient current waveforms. The model has been compared to experimental data where qualitatively full current waveform has been reproduced. The capability of the proposed model has been shown to provide even quantitative agreement within limitations given mainly by diffusion and electron repulsion. We have simplified the model even more based on our experience from the numerical simulations. The simplified analytical considerations have been shown to provide a fruitful insight into detector operation. The current waveforms were studied under varied doping density, positive and negative space charge, cathode and intergrid bias, diffusion length and detector shielding. The trends on how current waveforms are influenced by these parameters provide numerically insight on a detector optimization. A mutual influence of these parameters has been discussed and their possible contradiction was pointed out. Finally, we have quantified increased efficiency of the commonly employed differential sensing in hard X-ray and $\gamma$-ray detectors. The more elaborate extension of the work has been discussed in terms of nonlinear set of coupled drift-diffusion, continuity and Poissons equation.  
The presented work will be useful for quick insight on a radiation detector operation and optimization for even interdisciplinary scientific community involving broader physics and medical physics.  

\section{Acknowledgement}
Financial support from the Technological Agency of the Czech Republic under project TE01020445, and the Grant Agency of the Czech Republic under contract P102/16/23165S are gratefully acknowledged.

\end{document}